\documentclass[printer]{aa}
\usepackage{txfonts}
\usepackage[dvips]{graphicx}
\usepackage{color}
\usepackage{natbib}
\begin{document}

\title{Particle kinetic analysis of a polar jet from SECCHI COR data}

\author{L. Feng\inst{1,2}
\and B. Inhester\inst{2}
\and J. de Patoul\inst{2}
\and T. Wiegelmann\inst{2}
\and W.Q. Gan\inst{1}}
\institute{Key Laboratory of Dark Matter and Space Astronomy, Purple Mountain Observatory,
  Chinese Academy of Sciences, 210008, Nanjing, China
\and Max-Planck-Institut f\"ur Sonnensystemforschung,
       Max-Planck-Str.2, 37191, Katlenburg-Lindau, Germany\\
\email{lfeng@pmo.ac.cn}}

\date{Received / Accepted}

\abstract {}{We analyze coronagraph observations of a polar
jet observed by the Sun Earth Connection Coronal and Heliospheric
Investigation (SECCHI) instrument suite onboard the Solar
TErrestrial RElations Observatory (STEREO) spacecraft.}
{In our analysis we compare the brightness distribution of the jet
in white-light coronagraph images with a dedicated kinetic particle model.
We obtain a consistent estimate of the time that the jet was launched from
the solar surface and an approximate initial velocity distribution in
the jet source. The method also allows us to check the consistency of the
kinetic model.
In this first application, we consider only gravity as the dominant
force on the jet particles along the magnetic field.}
{We find that the kinetic model explains the observed brightness
evolution well. The derived initiation time is consistent with the
jet observations by the EUVI telescope at various wavelengths.
The initial particle velocity distribution is fitted
by Maxwellian distributions and we find deviations of the 
high energy tail from the Maxwellian distributions.
We estimate the jet's total electron content
to have a mass between $3.2\times10^{14}$ and $1.8\times10^{15}$ g. 
Mapping the integrated particle number along the jet trajectory to its
source region and assuming a typical source region size, we
obtain an initial electron density between $8\times10^9$
and $5\times10^{10}$ cm$^{-3}$ that is characteristic for the lower
corona or the upper chromosphere. The total kinetic energy 
of all particles in the jet source region amounts from 
$2.1\times10^{28}$ to $2.4\times10^{29}$ erg.}{}

\maketitle 

\section{Introduction}
Polar coronal jets were originally observed by the Extreme-Ultraviolet Imaging
Telescope (EIT) onboard the Solar and Heliospheric Observatory (SOHO)~
\citep{Domingo:etal:1995}. In white light, the polar jets discovered by LASCO/SOHO
\citep{StCyr:etal:1997} appear narrow and collimated, and expand rapidly as
they travel through polar regions. They are often associated with an 
Extreme-Ultraviolet (EUV) jet seen near the solar surface \citep{Wang:etal:1998}. These jets are often rooted
in bright, low-lying loop features and are similar in appearance to Soft X-ray (SXR)
jets \citep{Shibata:etal:1992,Moses:etal:1997}. The essential
acceleration mechanism for all these jets is very likely provided by
magnetic reconnection. The difference between the above and other
jet-like features, e.g. chromospheric jets, is the altitude where the magnetic
reconnection is assumed to occur. The higher energy jets tend to be
accelerated at a higher altitude than the lower energy jets
\citep{Shibata:etal:2007}. The EUV and SXR jets are often caused by the
reconnection in the upper chromosphere or the lower corona.

\citet{Wang:etal:1998} analyzed 27 jets in EIT and LASCO data and
characterized their motion by three different velocities: the leading-edge
velocity $v_{lead}$, the centroid velocity $v_{cen}$ and the initial
velocity $v_{init}$ of the centroid. In all cases
$v_{cen}$ was much less than $v_{lead}$, indicating that the jets stretched out
rapidly as they propagated through the corona. The authors also found that the bulk
of the jet material decelerated as it propagated from the limb to the C2 field
of view (FOV). This deceleration
was attributed to solar gravity. However, the combined results of
$v_{init} < v_{escape}$ and $v_{lead} > v_{escape}$, and the lack of evidence
for downflow in EIT and C2 led Wang et al. to propose some in situ
acceleration that prevents the bulk of the jet material from falling back onto the
Sun after it was ejected.

\citet{Wood:etal:1999} improved the jet velocity estimates by
\citet{Wang:etal:1998} using on height-time plots. The authors
determined the trace of the jet centroid from the EIT to the C2 FOV and
found that the observed kinematic trajectories could be fitted with some
success by ballistic orbits. They concluded, however, that gravity
alone was not the only force controlling the jet propagation. Because of the
similar behavior of the jets studied, both \citet{Wood:etal:1999}
and \citet{Wang:etal:1998} suggested that by the time jets reached the C2
FOV, they were incorporated into the ambient solar wind.

More recently, \citet{Ko:etal:2005} studied a jet observed jointly by 
several instruments above
the limb. These authors found that a ballistic model could explain most of the
dynamical properties of this jet. In their model it was assumed that the gas was
ejected upward from the surface with a range of initial speeds. The smooth
change of the upflow-to-downflow speed at a certain altitude derived from the
ballistic model was found to be consistent with the change of the line intensities
from Doppler dimming observed by UVCS/SOHO.
Owing to a lack of high cadence coronagraph observations Ko et al.'s study
was essentially confined to heights below 1.64 $r_\odot$.

On June 7, 2007 a big eruptive jet was observed by EUVI, COR1, and COR2 on
board the STEREO mission with higher spatial and temporal resolution compared
to the data from EIT and LASCO C2. It extended from the solar surface to
5~$r_{\odot}$. The event was also studied by \citet{Patsourakos:etal:2008} from
the stereoscopic viewpoint. They estimated the jet positions and the speed
of the leading front at different times in the EUVI FOV.

In this paper we will attempt to analyze the same jet based on white-light
coronagraph observations at heights beyond about 1.5 $r_\odot$. We extend
the ballistic approach by \citet{Ko:etal:2005} by quantitatively
comparing the density variation from Thomson scattered white-light brightness at
different heights to the variation expected from a ballistic model of the
jet particles.
This method has the advantage that it avoids the estimates of
jet centroids and fronts. These fronts are not well-defined for a diffusively
spreading plasma cloud, the jet centroids are often difficult to determine
because a substantial part of jet material is hidden behind the occulter.
Another advantage of our particle kinetic analysis lies is that it
also provides a test of the validity of the ballistic model.
We therefore receive much more information than the conventional leading edge,
centroid velocity measurements.
After the description of the observations in \S2, we will introduce the
ballistic model in \S3. In \S4 we present the results and try to
extrapolate our findings to the jet source region and also
discuss the limits of our model. Finally, we summarize our
conclusions in the last section.

\section{The data}

\subsection{The polar jet in EUVI, COR1, and COR2 images}
The jet we investigated was observed over a radial range from the solar
surface out to five solar radii. We used observations from the EUVI and
two white light coronagraphs (COR1 and COR2). They belong to the SECCHI
instrument suit \citep{Howard:etal:2008}. EUVI is a full disk imager with a
FOV of 1.7~$r_{\odot}$. COR1 and COR2 are two traditional Lyot
coronagraphs, with the FOV in the range of 1.5~$r_{\odot}$ to 4~$r_{\odot}$,
and 2.5~$r_{\odot}$ to 15~$r_{\odot}$, respectively. 
The angular resolution of one pixel in EUVI, COR1, and COR2 is 1.6 arcsec, 
7.5 arcsec, and 14 arcsec, respectively.

\begin{figure} % fig 1
  \frame{
  \vbox{
  \includegraphics[width=9cm, height=2cm]{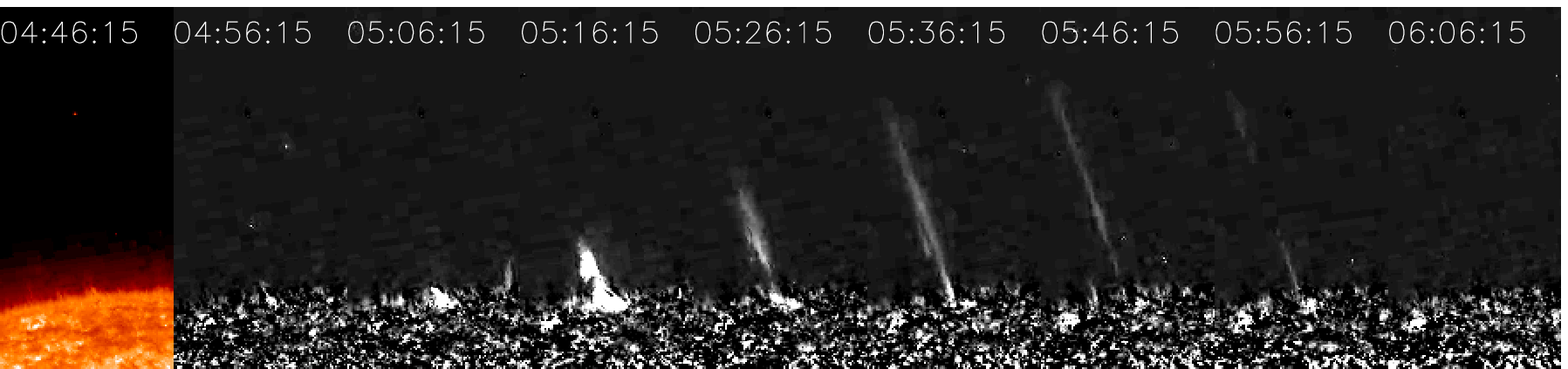}
  \hspace{0.1cm}
  \includegraphics[width=9cm, height=2cm]{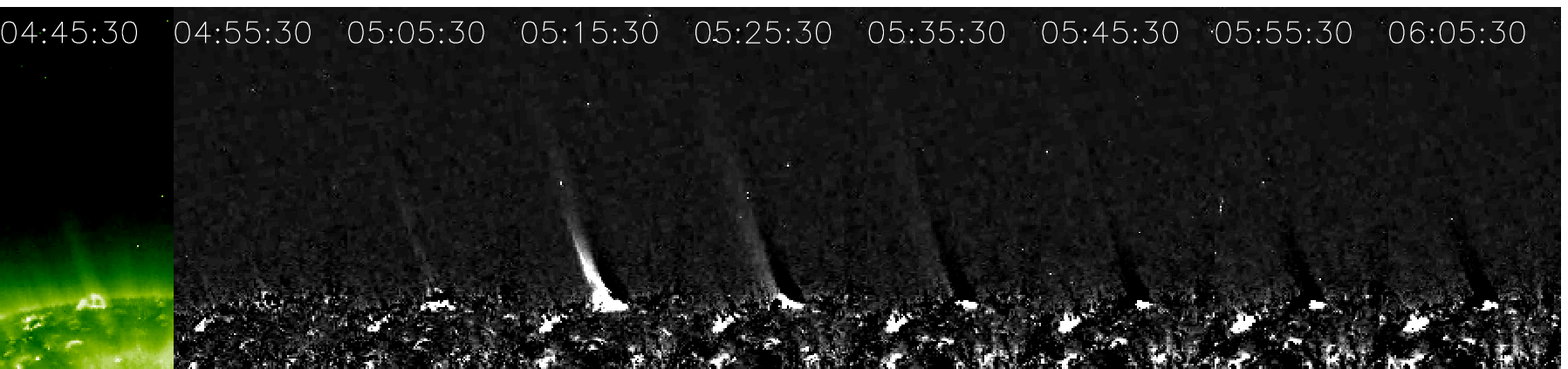}
  \hspace{0.1cm}
  \includegraphics[width=9cm, height=2cm]{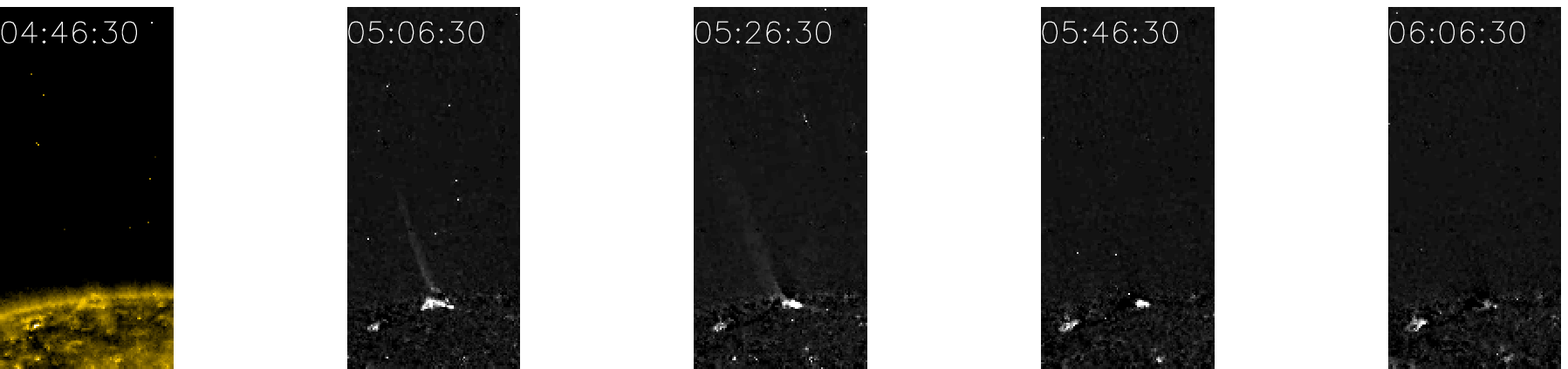}}}
  \caption{The jet time series of difference images at 304~{\AA},
  195~{\AA} and 284~{\AA} from top to bottom observed by the EUVI instrument
  on board STEREO A.}
  \label{fig:eu_304195284}
\end{figure}

\begin{figure*} % fig 2
  \includegraphics[width=15cm, height=3cm]{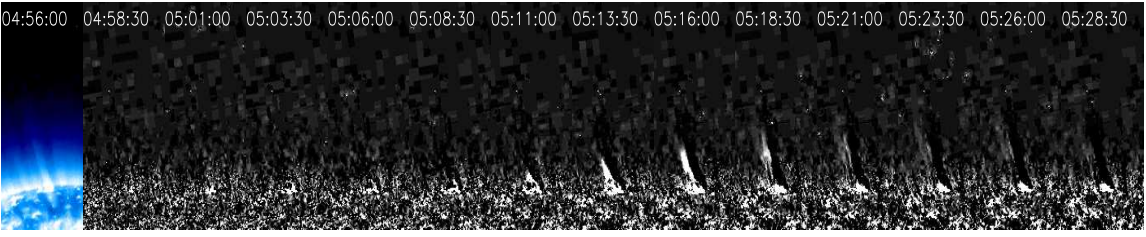}
  \caption{Jet time series of difference images at 171~{\AA} from
  EUVI A. The FOV is $320^{\prime\prime}\times672^{\prime\prime}$.}
  \label{fig:eu_171}
\end{figure*}

In EUV, the jet could be observed at all four
wavelengths from the EUVI telescope. The time cadence was 10 minutes for
304~{\AA} and 195~{\AA}, and 20 minutes for 284~{\AA}. For
171~{\AA}, the temporal resolution was as high as 2.5 minutes. In
Figs.~\ref{fig:eu_304195284} and \ref{fig:eu_171} we show the difference
images of the event at four wavelengths of EUVI observed by STEREO A. They
were created by
subtracting the pre-event images shown in color in the leftmost column of
Figs.~\ref{fig:eu_304195284} and \ref{fig:eu_171}. The following frames
displaying the jet at later times is running from left to right with time.
Nearly simultaneous images at different wavelengths are stacked vertically.

The EUVI observations indicate that this jet contains both
hot and cool material. The jet eruption first clearly appeared in hotter lines
195~{\AA} and 284~{\AA} at around 05:06 UT, later in 171~{\AA} at around 05:11,
and finally in 304~{\AA} at around 05:16 UT. The jet life time appeared to be
shorter in the hotter lines (171, 195 and 284~{\AA}), and was clearly longer
in the cooler line~(304~{\AA}). More detailed descriptions of the EUV
observations of this jet are provided in \citet{Patsourakos:etal:2008}.

%\begin{figure*} % fig 1
%  \frame{
%  \vbox{
%  \includegraphics[width=13.75cm, height=3cm]{eu304.eps}
%  \hspace{0.1cm}
%  \includegraphics[width=13.75cm, height=3cm]{eu195.eps}
%  \hspace{0.1cm}
%  \includegraphics[width=13.75cm, height=3cm]{eu284.eps}}}
%   \caption{Time series of difference images 
%   observed by the EUVI instrument onboard STEREO A on June 7, 2007. The first 
%   row shows the images at 304~{\AA},
%  the second row at 195~{\AA} and the last row at 284~{\AA}.
%  The FOV of each frame is $320^{\prime\prime}\times672^{\prime\prime}$.}
%  \label{fig:eu_304195284}
%\end{figure*}

\begin{figure*} % fig 3
  \vbox{
  \includegraphics[width=15cm, height=3cm]{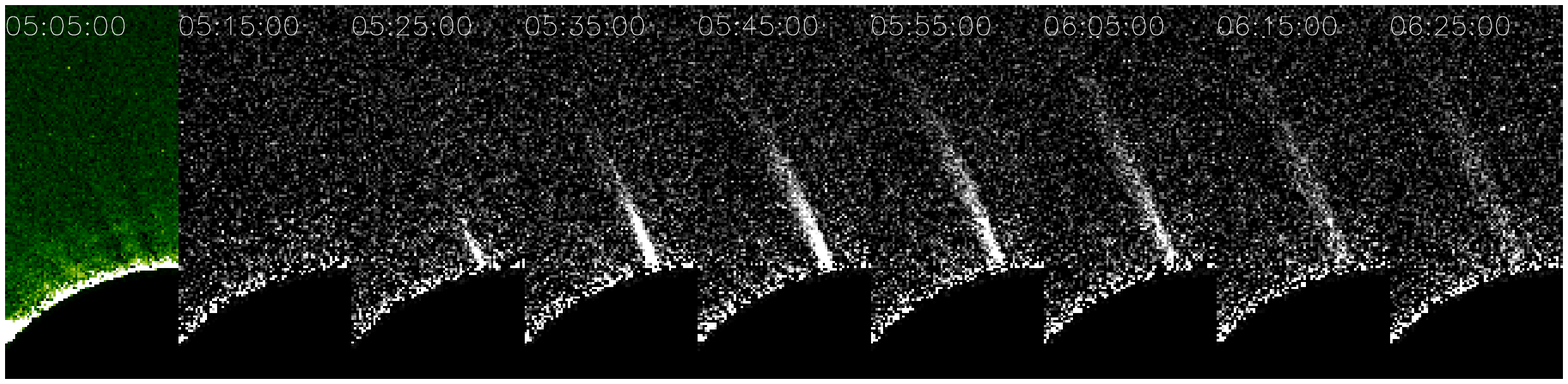}
  \includegraphics[width=13.3cm, height=3cm]{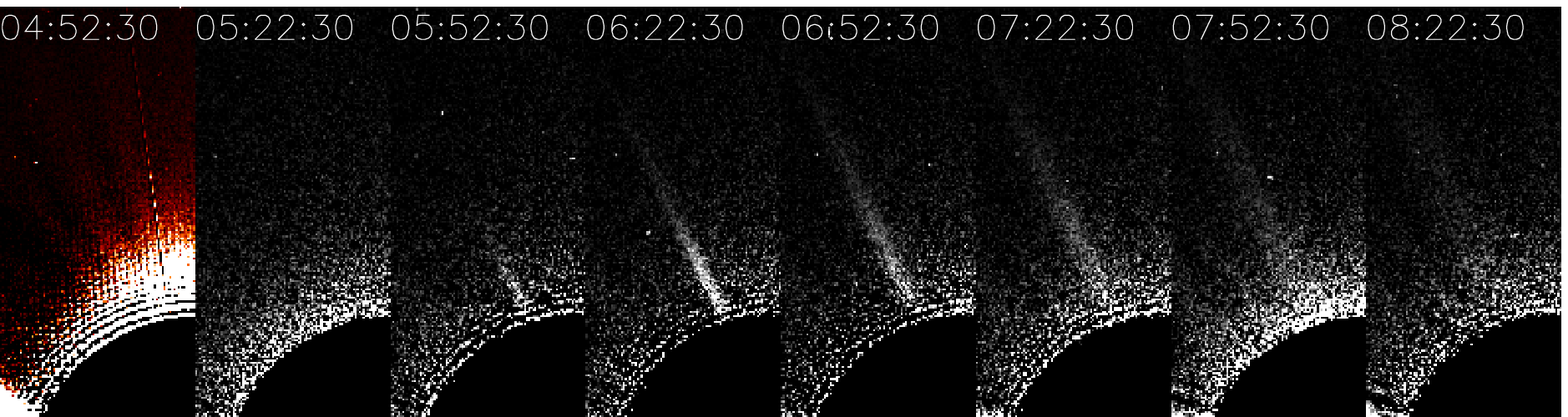}}
  \caption{Jet time series of difference total brightness images
  observed by COR1 (upper row) and COR2 (bottom row) onboard STEREO A. The FOV
  of each frame in the upper row is $1065^{\prime\prime}\times2355^{\prime\prime}$,
  in the bottom row $2744^{\prime\prime}\times5614^{\prime\prime}$. As a reference,
  the solar radius was around $1002^{\prime\prime}$ as observed by STEREO A.}
  \label{fig:co12}
\end{figure*}

\begin{figure} % fig 4
  \includegraphics[width=8cm, height=7.5cm]{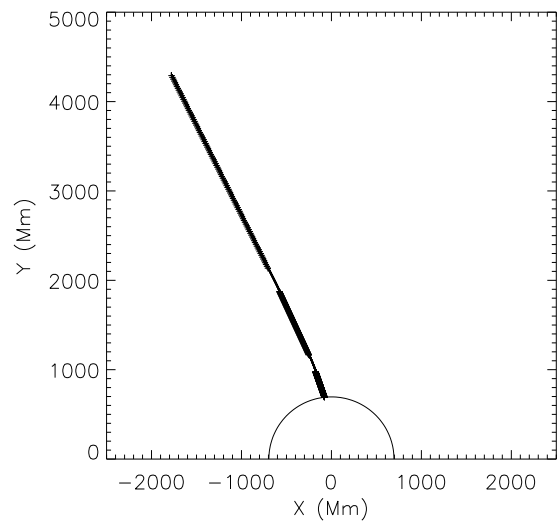}
  \caption{Jet geometry in units of Mm trace from EUVI to COR1, then to
  COR2 (thick curve segments). The thin solid curve is a spline fit to the three
  traced segments. The half circle indicates the solar limb.}
  \label{fig:euco12}
\end{figure}

The jet we are studying here was also traced in the COR1 and even in the
COR2 field of view. These observations are the main object of our analysis. 
COR has a polarizer measuring polarized brightness in three
directions separated by $120 ^\circ$. From this the total brightness is obtained
by the standard procedure $secchi\_prep$. In
Fig.~\ref{fig:co12} we show the total brightness of the jet observations as
difference images with respect to the closest pre-event image. Again the
images are arranged with observation time from left to right and stacked
roughly synchronously for different telescopes.

In Fig~\ref{fig:euco12} the jet geometry traced in the FOV of EUVI,
COR1, and COR2 is shown by the three thick curve segments. A smoothing spline
is employed to connect the jet orbit over the entire height range. In the next
section the trajectory of jet particles is calculated along this thin solid
curve in Fig~\ref{fig:euco12}. The good alignment of the jet segments from the
three instruments indicates that they are part of the same jet. Notice that
there was a slight change of about 6 degrees between the jet direction close
to the surface as seen in EUVI and at altitudes beyond $1.5~r_\odot$ seen in
COR1.
As a summary, the jet extended out to 5~$r_{\odot}$. The lifetime in EUVI was
around several tens of minutes, in COR1 the event could be observed for
about 1.5 hours and in COR2 for about 3 hours.

At 171~{\AA} which is the preferred EUV wavelength for plume observations,
we noticed a preexisting plume to the right-hand side of the jet.
In COR1, the plume was found cospatial with the jet.
It is not exceptional that a plume is aligned with a jet in a projected
2D image \citep{Wilhelm:etal:2011}. However, this more or less close alignment
along the same line of sight (LOS) may be a coincidence.
In STEREO B images at 171~{\AA} we also found a preexisting plume slightly to
right-hand side of the jet, in COR1 B the plume signal was too faint to conclude
that the plume was really close to the jet. In view of the small separation of
the two spacecraft of only 10 degrees during this event, we still hesitate to
assume that the jet has a physical relation to the preexisting plume.
By using difference images, the emission of the plume is eliminated
for the subsequent analysis.

\subsection{The distance-time brightness relation}

\begin{figure} % fig 5
  \includegraphics[width=9cm, height=10cm]{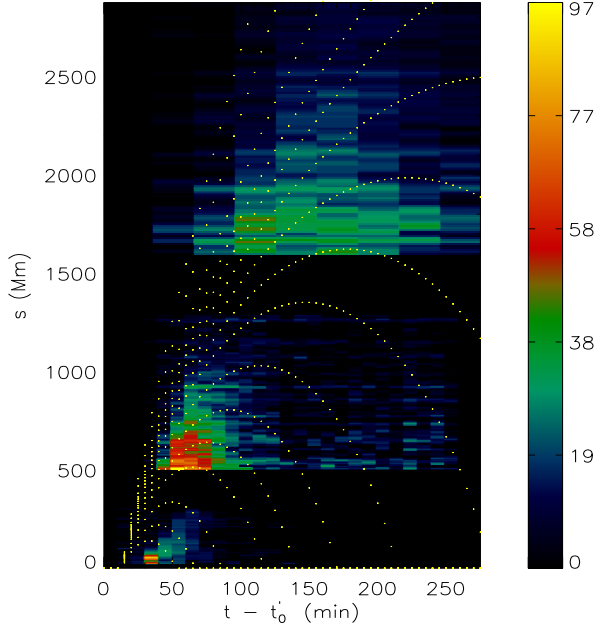}
  \caption{Image intensity for EUVI 304 and brightness for COR1 and
  COR2 $B(s,t-t\,^{'}_0)$ as a function of time and distance along the jet axis. 
  $t\,^{'}_0 =$ 04:46 UT, which is a reference time.
  The intensities and brightnesses were integrated across the
  jet width from the difference images (Figs.\ref{fig:eu_304195284} 
  and \ref{fig:co12}). The color code on the right indicates 
  the 304 intensity in units of number of photons. From bottom to top 
  the yellow dotted lines indicate particle trajectories with
  different initial velocities of 250, 300, 350, 400, 425, 450,
  475, 500, 515, 530, 545, 560, 575, 600, 650, and 700 km\,s$^{-1}$.}
  \label{fig:DT_euco}
\end{figure}

Based on the difference images above, we determined the jet intensity in the
EUVI images at 304~{\AA} and the white-light total brightness in COR1 and
COR2 images for successive times and different distances along the jet axis.
The total brightness was integrated across the jet cross section for
each exposure and for each position along the jet axis 
after the pre-event background was subtracted. By integrating 
the brightness across the jet width, at each distance $s$ the density decrease 
caused by the magnetic field line divergence was removed. Note that the
integration along the line-of-sight direction is
implicit in the observation of the optically thin jet plasma.
The distance $s$ along
the jet axis remains the only relevant spatial coordinate.
The resulting distance-time (DT) total brightness plot is shown in
Fig.~\ref{fig:DT_euco}.
For a clearer view, the brightnesses of COR1 and COR2 in
Fig.~\ref{fig:DT_euco} are multiplied by $10^9$ and $10^{10}$, respectively, 
to match the
intensity at 304~{\AA}. Therefore the absolute values in
Fig.~\ref{fig:DT_euco} are not comparable among different instruments.
The distance on the ordinate is the length along the jet axis from its
footpoint at 1~$r_{\odot}$.
The sampling $ds$ in the direction along the jet axis corresponds to the size
of one pixel in the respective original images.
The time is in units of minutes and starts at 
$t\,^{'}_{0} =$ 04:46 UT, which was the earliest
observation time of the jet in all frames in Figs.~\ref{fig:eu_304195284},
\ref{fig:eu_171}, and \ref{fig:co12}. It ends at
09:21 UT when we were barely able to see a jet signal any more.

The white-light signal in COR1 and COR 2 is caused by Thomson scattering
at free electrons and therefore the total brightness observed by COR1 and COR2
is proportional to the line-of-sight integrated column density of coronal
electrons.
Since in Fig.~\ref{fig:DT_euco} we have subtracted the background and
integrated the coronagraph signal across the jet cross section, the
resulting image pixel count $B(s,t-t\,^{'}_0)\,ds$ for COR1 and COR2
in Fig.~\ref{fig:DT_euco} is proportional to the number of jet particles
at a distance $s$ along the jet in a height range $ds$ resolved by the
image pixel, i.e.,
\begin{equation}
  B(s,t-t\,^{'}_0)\,ds \propto N(s,t-t\,^{'}_0)\,ds,
  \label{equ:jacob_B1}
\end{equation}
where $N(s,t-t\,^{'}_0)ds$ is the number of jet particles in a range $ds$
at a distance $s$ from the foot point of the jet and at a time $t$.
$B(s,t-t\,^{'}_0)$ is brightness of the corresponding pixel scaled for the
different pixel sizes of COR1 and COR2.

\section{Ballistic model}

For the analysis below, we assumed that in the investigated jet 
a package of particles ejected upwards with different speeds $v_0$ simultaneously 
during the initiation process.  
Note that parameter $v_0$ is the velocity component parallel to the jet axis. We did
not consider the details of the perpendicular velocity, which is related to
the gyro motion of particles. Moreover, all particles that move upward follow
the jet axis traced in the images from STEREO A. Because we observed the jet very close to the 
solar limb, we assume that the projection effect does not effect
the results much.

\subsection{The particle trajectory along the jet}

The motion of a charged particle along the
unperturbed magnetic field is obtained after averaging the forces
over the particle gyro phase
\begin{equation}
\frac{d\tilde{s}}{dt}=\tilde{v}\;,\quad
\frac{d\tilde{v}}{dt}=
 -g_{\odot}\frac{r_{\odot}^2}{r(\tilde{s})^2}\cos{\alpha(\tilde{s})}
 - \frac{\mu}{m}\hat{s}\cdot\frac{\partial}{\partial s}\mathbf{B}(\tilde{s})
 - \mathbf{a}_\mathrm{coll}
 \label{equ:kinem1}
\end{equation}
 \[
 \tilde{s}(0,v_0)=r_\odot\; ,\quad
 \dot{\tilde{s}}(0,v_0)=v_0.
 \]
Here, $\tilde{s}(t-t_0,v_0)$ denotes the distance along the field line
and $\tilde{v}(t-t_0,v_0)$ the corresponding velocity 
in which $t_0$ is the jet initiation time.
The first term on the right-hand side is the gravity force at radius $r(s)$
from the Sun center, $g_{\odot}$ is the gravity on the solar surface
and $\alpha$ is the inclination of the local magnetic field
that measures the angle between the local radial direction and the 
tangent of the jet axis.
The second term represents the mirror force driven by the particle's
invariant magnetic moment $\mu = mv_\perp^2/2B$ and the final term accounts
for the deceleration of the jet particle by collisions with the
background plasma. In the following we ignore the mirror and
collision term because they are small at heights above 1.5 $r_\odot$
where the coronagraph observations were made.
In Sect.~\ref{sec:mirror+coll} we justify this choice in
more detail and discuss possible modifications from our results
concerning the jet properties closer to the surface.

We solved the second-order ordinary differential equation
(Equ.~\ref{equ:kinem1})numerically by a fourth-order Runge-Kutta method.
Each resulting orbit depends
on two parameters, the initial velocity $v_0$ and the time $t_0$ that a particle
is ejected from the surface. Because we assumed that the particles were
launched in a unique jet event, $t_0$ was chosen to be identical for all orbits.
Examples of the calculated trajectories for
different initial velocities are plotted as dots in
Fig.~\ref{fig:DT_euco} with $v_0$ in the range from 250 to 700
km\,s$^{-1}$. The orbit whose apex is close to the edge of the COR1
occulter at a distance of 500 Mm from the surface has the initial velocity
400 km\,s$^{-1}$.

We have overplotted some particle trajectories in Fig.~\ref{fig:DT_euco}
for a somewhat arbitrarily chosen starting time $t_0$. 
Qualitatively, we find that the brightness of a pixel in the DT-diagram
is roughly proportional to the number of orbits across it.
Because the brightness in white-light coronagraphs
is caused by the Thomson scattering and is hence proportional to the electron
density, the variation in the jet brightness will be
controlled by the particle motion.
The distance between particles starting at the same time with different
initial velocities will grow continuously according to their
orbit $\tilde{s}(t-t_0,v_0)$ obtained from Equ.~\ref{equ:kinem1}. To the extent that the
particle position will disperse, the observed brightness of the jet will
decrease. This scenario qualitatively agrees with the observations in
Fig.~\ref{fig:DT_euco}.
In the next section, we quantify this phenomenon.

\subsection{The Jacobian}

\begin{figure} % fig 6
  \includegraphics[width=8cm, height=7.5cm]{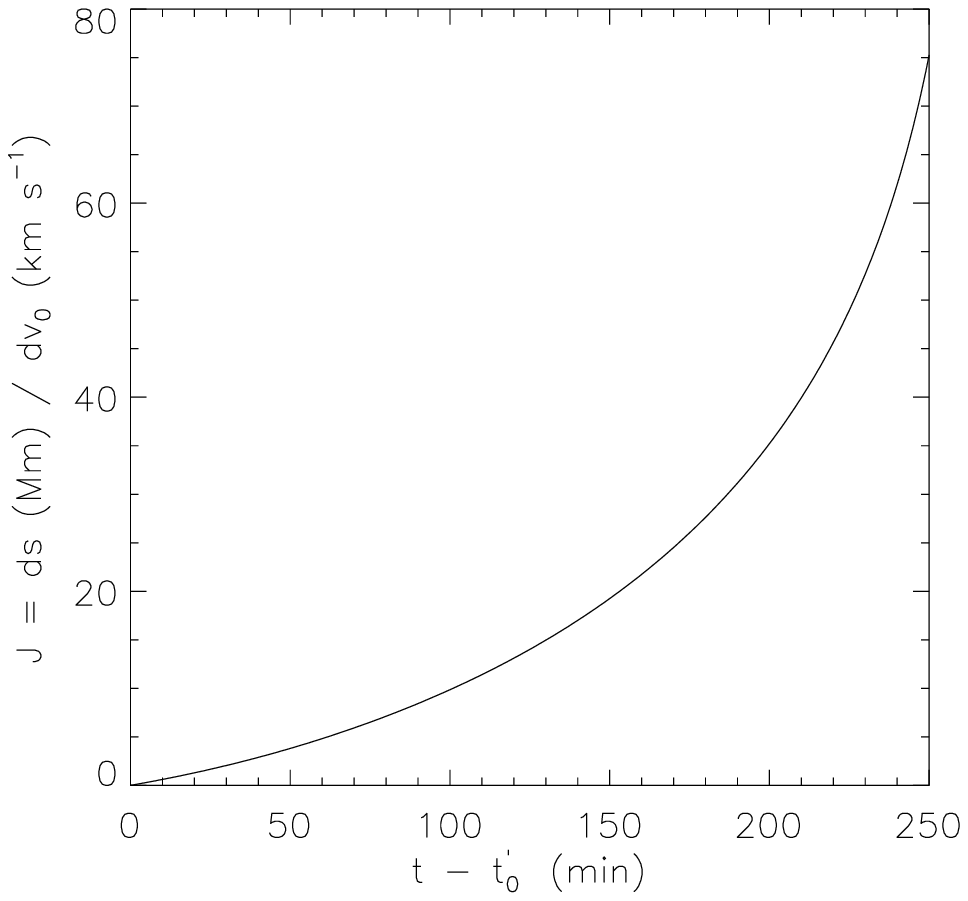}
  \caption{One example of the Jacobian as a function of time for
  $v_0 = 500$ km\,s$^{-1}$ and $t_0 = t\,^{'}_0 =$ 04:46 UT. $t_0$ is the
  jet initiation time and $t\,^{'}_0$ is the reference time at 04:46 UT.}
  \label{fig:jacob}
\end{figure}

To quantitatively compare the brightness variation caused by the ballistic
particle motion to the observations from white light coronagraphs COR1 and
COR2, we estimated the jet particle density (Equ.~\ref{equ:jacob_B1}) 
from the particle trajectories (Equ.~\ref{equ:kinem1}). For this purpose, 
we equated the number of jet particles $N(s,t-t\,^{'}_0)$ in distance range $s$ to $s+ds$
at time $t$ to the number of particles that at time $t_0$
were just in the right range of initial velocities $v_0$ to $v_0+dv_0$
to reach the height range $s$ to $s+ds$ at time $t$.
\begin{equation}
N(s,t-t\,^{'}_0)\,ds=f(v_0)\,dv_0
  \label{equ:jacob_B0}
\end{equation}
Here, $f(v_0)$ is the initial velocity distribution and 
$s=\tilde{s}(t-t_0,v_0)$ depends on the initial particle velocity.
Combining Equ.~\ref{equ:jacob_B1} and \ref{equ:jacob_B0}, we arrive at
\begin{eqnarray}
B(s,t-t\,^{'}_0)\propto N(s,t-t\,^{'}_0)& = & \frac{f(v_0)}{J(t-t_0,v_0)}
\;, \\
  \label{equ:jacob_B2}
\quad\mathrm{where} \quad J(t-t_0,v_0)& = & \frac{d\tilde{s}(t-t_0,v_0)}{dv_0}
  \label{equ:jacob_B3}
\end{eqnarray}
%\begin{equation}
% B(s,t-t\,^{'}_0)\propto N(s,t-t\,^{'}_0) =  \frac{f(v_0)}{J(t-t_0,v_0)}
% \;,\quad\mathrm{where} \quad J(t-t_0,v_0) =  \frac{d\tilde{s}(t-t_0,v_0)}{dv_0}
%\label{equ:jacob_B2}
%\end{equation}
is the Jacobian of the particle orbit. It quantifies the
increasing dilution of the particle density along the jet with time.
In Fig.~\ref{fig:jacob}, an example of $J(t-t_0,v_0)$ is plotted vs time for
$v_0 = 500$ km\,s$^{-1}$. If initially $v_0$ deviates slightly, say 
$dv_0 = 1$ km\,s$^{-1}$, after a certain time, say 50 mins, the difference in 
distance along the jet is about 4 Mm because Fig.~\ref{fig:jacob} shows 
that $J$ at $t -t_0 = 50$ min is about $4\times10^3$ s. 

\section{Results and discussions}

In this section, the observed brightness distribution $B(s,t-t\,^{'}_0)$ along the
particle trajectories with given initial velocities $v_0$ is fitted by
the corresponding Jacobians $J(t-t_0,v_0)$ to derive the scaling factors
$f(v_0)$ in Equ.~\ref{equ:jacob_B2} and the optimal jet initiation time $t_0$.

\subsection{Fit of the Jacobian to the brightness}

\begin{figure*} % fig 7
  \centering
  \vbox{\hbox{
  \includegraphics[width=6.3cm, height=5.5cm]{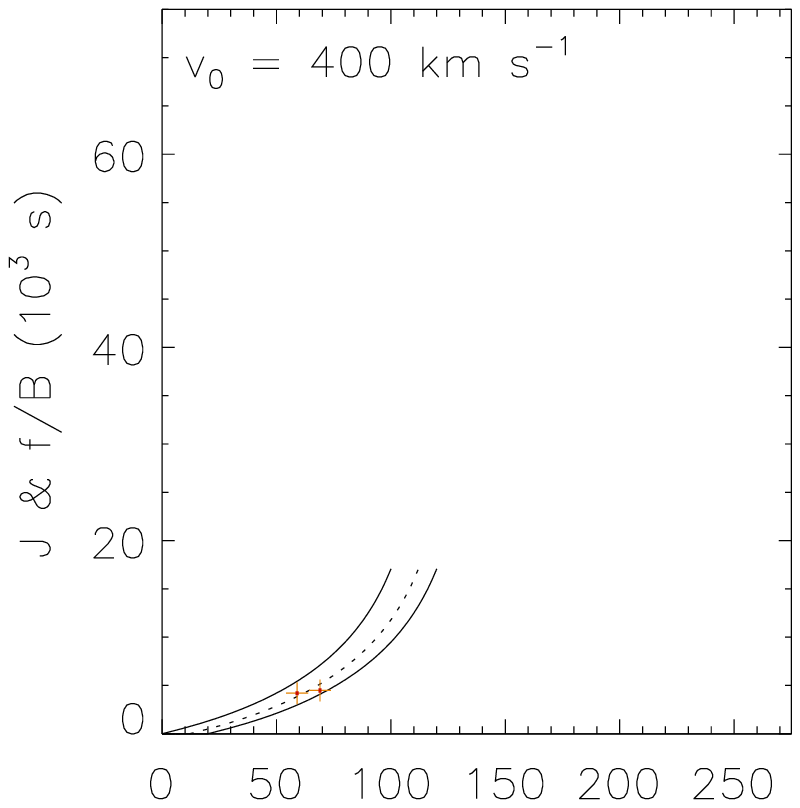}
  \includegraphics[width=6.3cm, height=5.5cm]{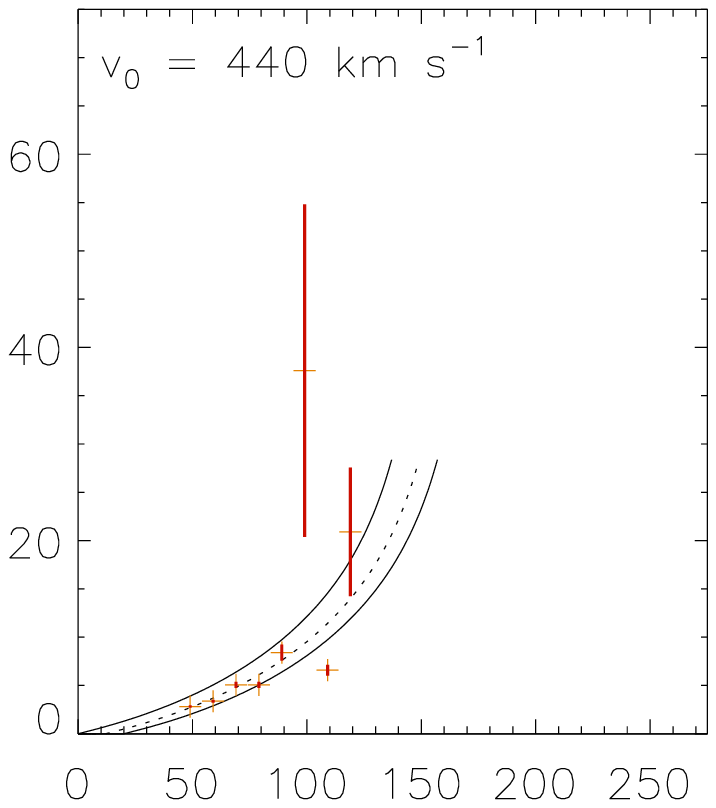}}
  \hbox{
  \includegraphics[width=6.3cm, height=5.5cm]{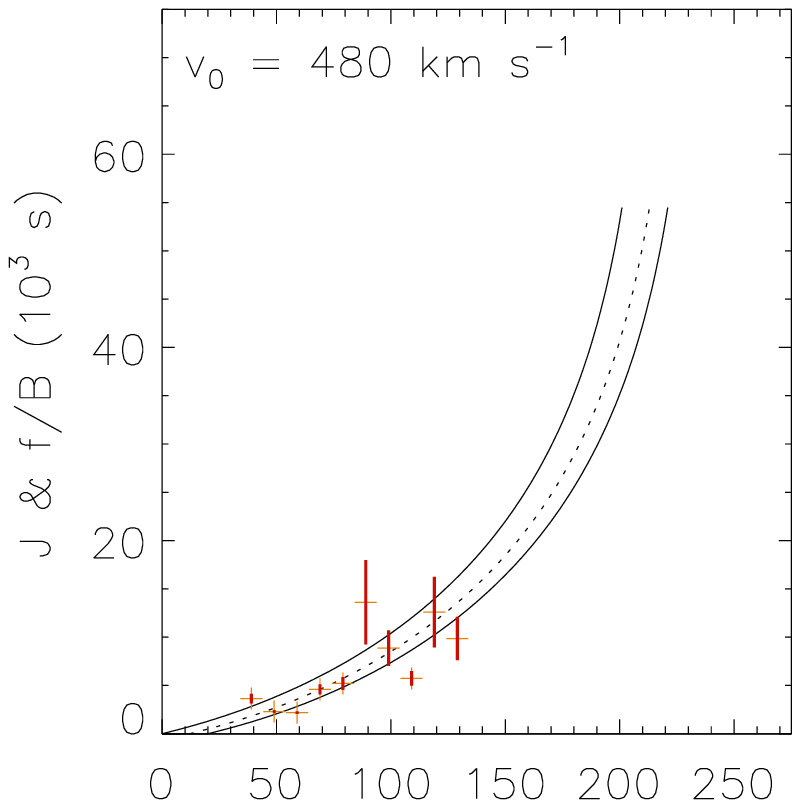}
  \includegraphics[width=6.3cm, height=5.5cm]{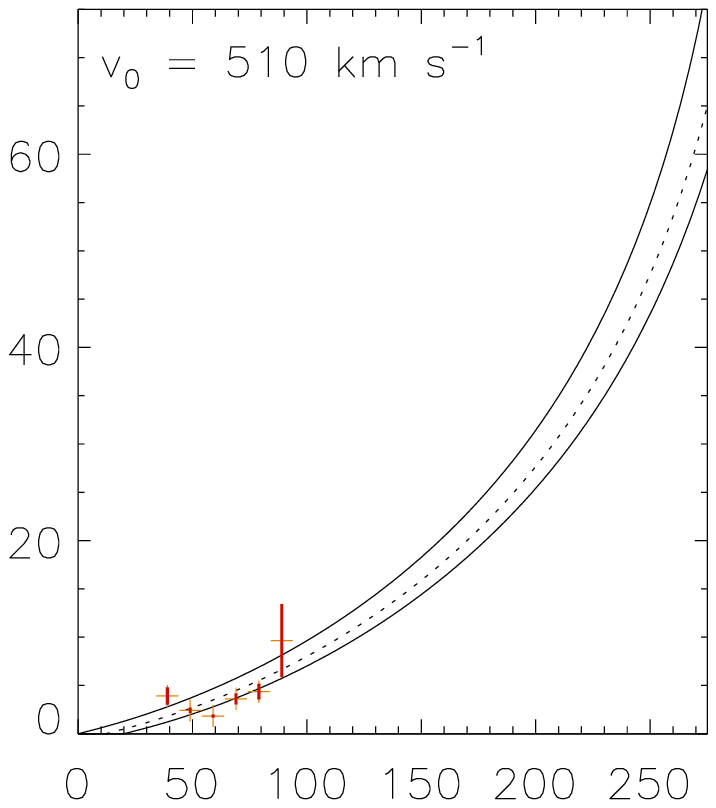}}
  \hbox{
  \includegraphics[width=6.3cm, height=5.5cm]{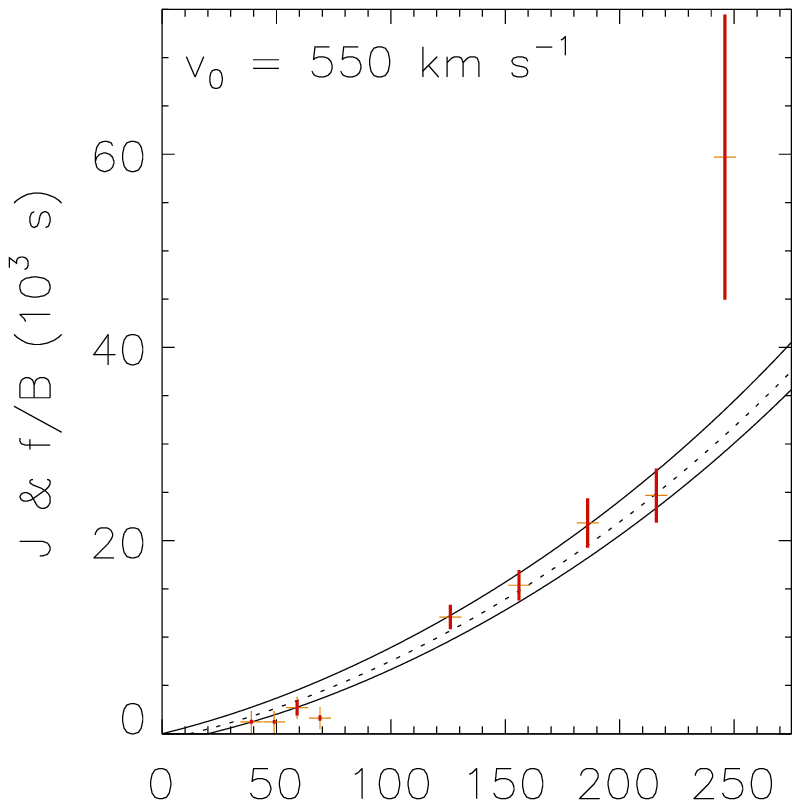}
  \includegraphics[width=6.3cm, height=5.5cm]{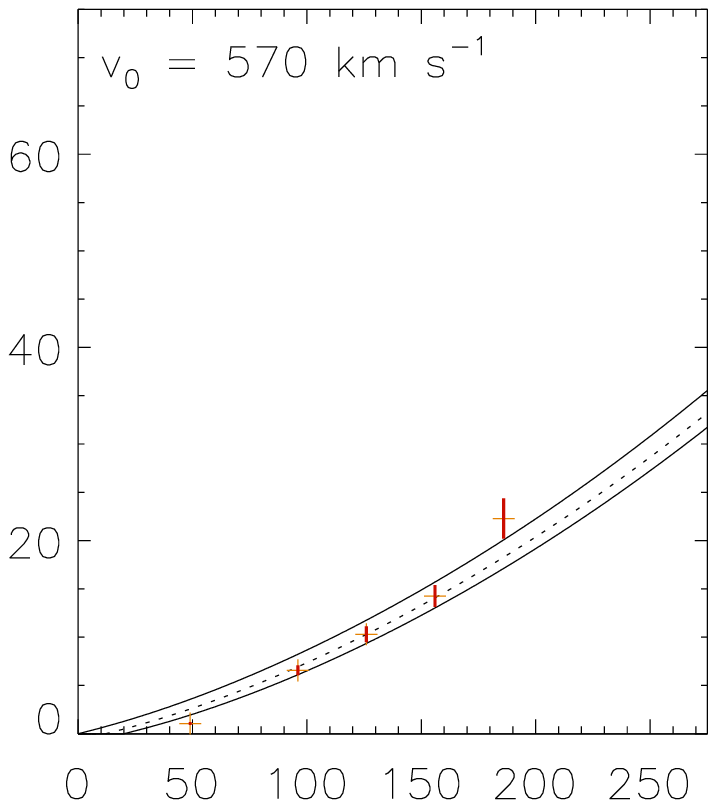}}
  \hbox{
  \includegraphics[width=6.3cm, height=5.5cm]{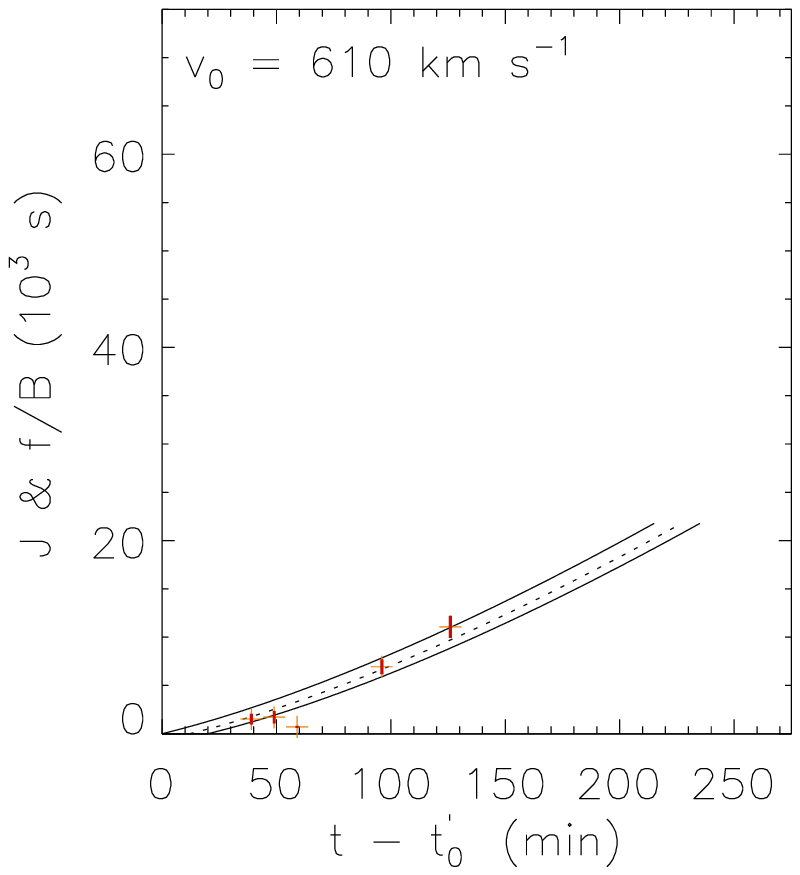}
  \includegraphics[width=6.3cm, height=5.5cm]{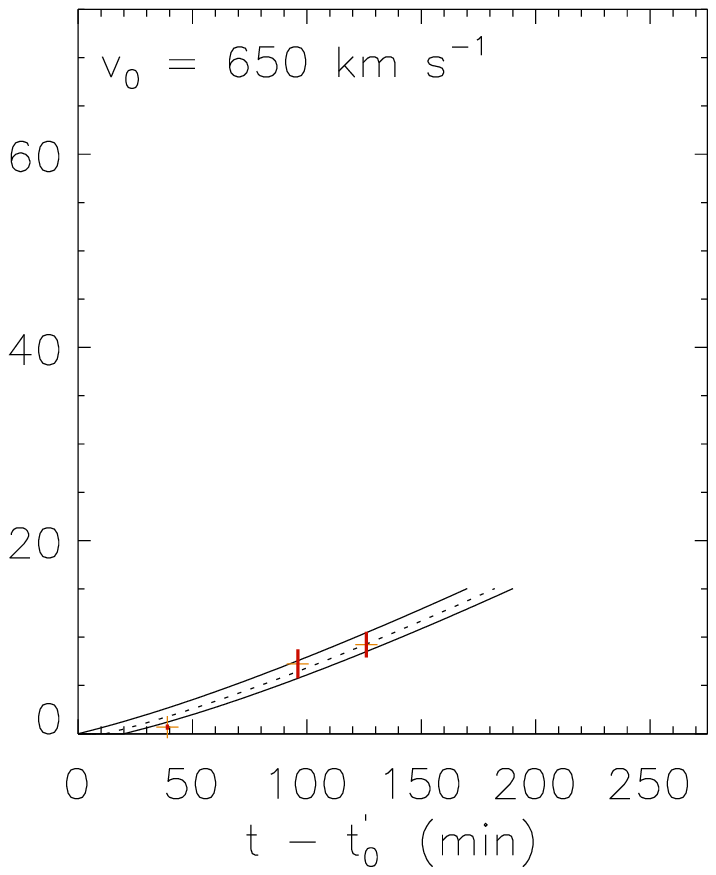}}}
  \caption{Comparison of the inverse of the observed brightness
  $B(s,t-t\,^{'}_0)$ times a fitted scaling factor $f(v_0)$ (red vertical
  bars) and the theoretical Jacobian $J(t-t_0,v_0)$ (black dashed
  curve) as a function of time $t-t\,^{'}_0$ after jet initiation at $t_0$.
  $t\,^{'}_0$ is a reference time at 04:46 UT.
  The different diagrams are obtained for different initial
  velocities $v_0$, which lead to different particle orbits
  $s=\tilde{s}(t-t\,^{'}_0, v_0)$ along which $B(s,t-t\,^{'}_0)$ is recorded. The
  vertical range of the red bars indicates the estimated error in $B$.
  The width between the two solid black curves on both sides of the
  dased curve shows the variation of $J$ due to an uncertainty in $t_0$.
  For more details see text.}
\label{fig:jf0Berr}
\end{figure*}

From Equ. 4, we expect the observed $B(s,t-t\,^{'}_0)$ to be proportional
to the inverse of the theoretical $J(t-t_0,v_0)$ for every $v_0$
if the correct orbit $s=\tilde{s}(t-t\,^{'}_0,v_0)$ is used in the first
argument of $B$ and the initial time $t_0$ is chosen correctly. The
constant of proportionality, $f(v_0)$, should then yield the
unscaled initial velocity distribution of the jet particles. Since
$J(0,v_0)=0$, we rather use the inverse of Equ. 4 to obtain
estimates for $t_0$ and $f(v_0)$ from fits of both sides of the
equation. In Fig.~\ref{fig:jf0Berr}, we show diagrams for different initial
velocities $v_0$ of the observed $1/B$ times the fitted scaling
factor $f$ (shown as red crosses) in comparison to the theoretical
$J$ (black dashed line) for our best estimates of $t_0$
and$f(v_0)$. The initial velocities are chosen to be in the range from
400 to 650 km\,s$^{-1}$.

The uncertainty in $J(t-t_0,v_0)$ owing to a possible error in $t_0$
is indicated by two solid black curves shifted with respect to the
central dashed curve by the uncertainty in the initiation time. The
estimated errors of $f/B$ along the respective trajectory
$\tilde{s}(t-t\,^{'}_0,v_0)$ are indicated by the vertical range of the
red bars. The error estimate was obtained from the noise
in the COR1 and COR2 difference images.
At each position $s$ along the jet, the noise was calculated
according to the standard deviation $\sigma$ of the brightness along a circle
centered at the Sun center with its radius reaching the position $s$.
In Fig.~\ref{fig:jf0Berr} only data points with a brightness higher than
$3\sigma$ are included. The uncertainty of $f/B$ indicated
by the red bar represents the $\pm3\sigma$ levels of the brightness $B$.

The slope of $J(t-t_0,v_0)$ at $t \rightarrow t_0$ is independent
of $v_0$. The enhanced slope of $J$ with time reflects the decrease of effective
gravity with the distance from the Sun. Particles with higher initial velocities 
tend to have less dilution of density (smaller slopes for $J$), which is indicated
qualitatively in Fig.~\ref{fig:DT_euco} as well. The plot of curve $J$
in Fig.~\ref{fig:jf0Berr} is terminated either at the time when
jet particles hit the solar surface in the cases of low $v_0$ or
at the time when the particles leave the FOV in Fig.~\ref{fig:DT_euco}
in the cases of high $v_0$.

According to Equ.~\ref{equ:jacob_B2}, there are two parameters that can be
modified to obtain a close fit between $J(t-t_0,v_0)$ and $f(v_0) / B(s,t-t_0)$.
One is the jet initiation time $t_0$ which is the same for all diagrams. A
change in $t_0$ corresponds to a horizontal shift of the red bars in all
diagrams simultaneously. Next, there is a scaling factor $f(v_0)$ for each
diagram individually. This factor represents the relative number of particles
with this initial velocity in the jet source.
Since Equ.~\ref{equ:jacob_B2} still includes a global proportionality
constant, the factors $f(v_0)$ represent only relative number densities at
this stage of our analysis. All parameters were optimized for a best fit.

From the EUV observations we conclude that the jet started between 04:46 UT and
05:06 UT. The first clear jet signal was seen at 05:06 UT in 284~\AA~and
195~\AA. There was a small brightening at 04:56 UT as well. However, the
signal was too weak to be identified as the jet initiation. Therefore we
assume that this jet was ejected in the time range from 04:46 UT to 05:06 UT.
This uncertainty is also represented graphically
in Fig.~\ref{fig:jf0Berr} by the two solid curves to either side of
the central dotted line representing the Jacobian.
An initiation time $t_0 =$ 04:58 UT gives the least sum of
the chi-squared deviations for all initial velocities $v_0$. 
We found a three minute uncertainty, which corresponds to the range of $t_0$ 
producing a $\chi^2$ enhancement by 5 \% above $\chi^2_{min}$. 

The comparison in Fig.~\ref{fig:jf0Berr} implies that the inverse
brightness follows the Jacobian curves quite closely. Although the fit is
not perfect, we can say that the ballistic model in general can explain
the particle kinematic behavior, and hence the brightness
variation in this jet.

\subsection{The jet source}

\begin{figure} % fig 8
  \vbox{
  \includegraphics[width=7.cm, height=6.5cm]{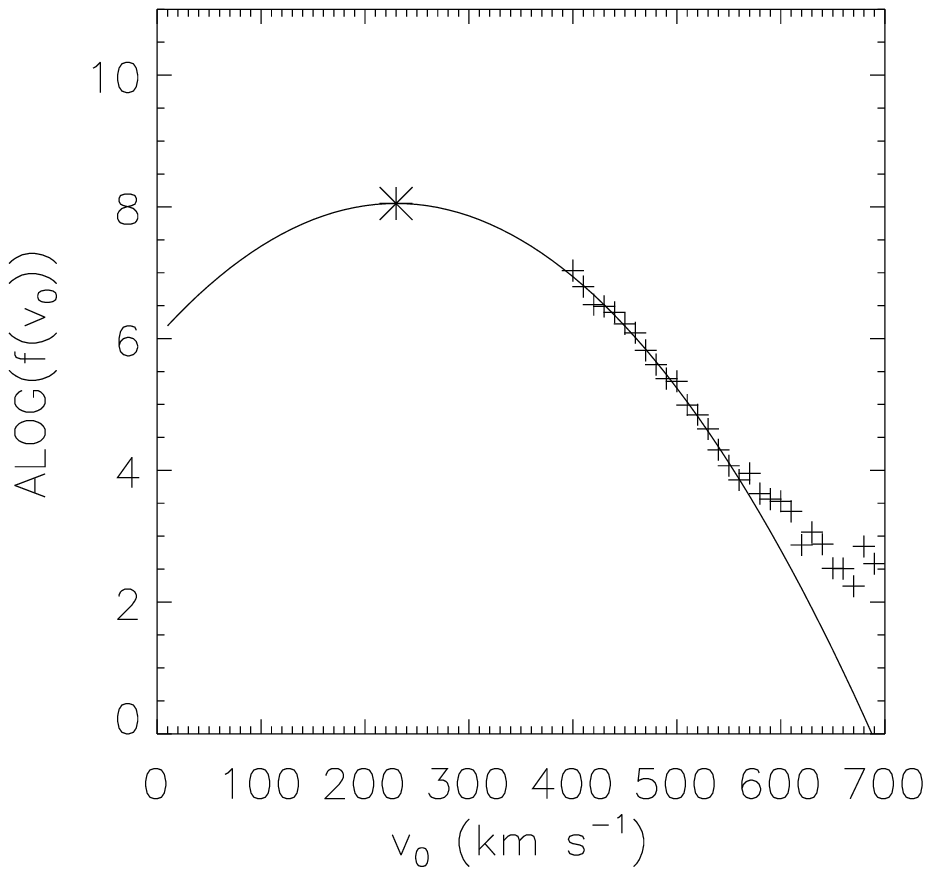}
  \includegraphics[width=7.cm, height=6.5cm]{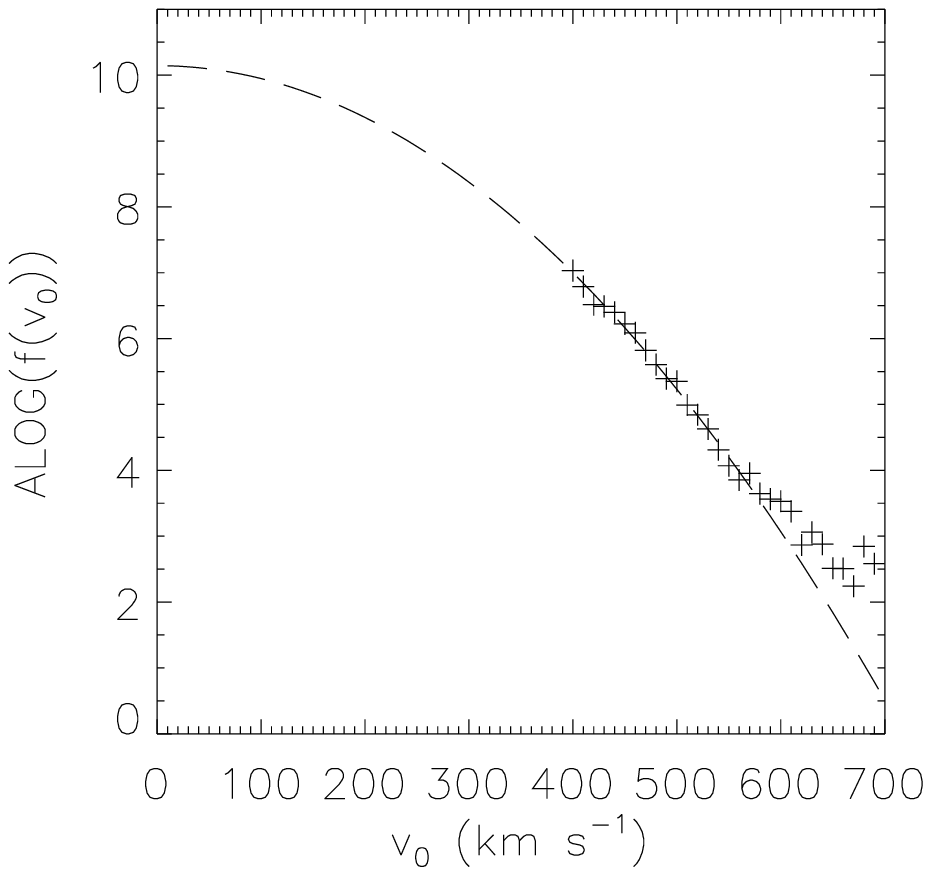}}
  \caption{Top: the distribution of the scaling factor $f(v_0)$ as a
  function of $v_0$ in natural logarithm with a fit by the Maxwellian
  distribution. Bottom: similar to the top one, the fit by the Maxwellian
  distribution with fixed $v_b = 0$ km\,s$^{-1}$.}
  \label{fig:f0v0}
\end{figure}

As mentioned above, for each $v_0$ the fit of $J(t-t_0,v_0)$ to $B(s,t-t\,^{'}_0)$
yields a scaling factor $f(v_0)$ that is proportional to the total number
of particles with this initial velocity.
Therefore, the distribution of $f(v_0)$ as a function of $v_0$ contains
information about the initial velocity distribution in the jet source region.
This velocity distribution $f(v_0)$ is shown in Fig.~\ref{fig:f0v0}
together with a fitted Maxwellian distribution
\begin{equation}
f(v_0)=C\sqrt{\frac{m}{2\pi \epsilon_k}}exp\left[\frac{-m(v_0-v_b)^2}{2\epsilon_k}\right].
\label{equ:maxw1d}
\end{equation}
Here $m$ is the averaged atomic mass in the solar corona and approximated as
$1.27m_p$, $k$ is the Boltzmann constant, $\epsilon_k$ is the 
mean kinetic energy of the particles in the source region
and $v_b$ a velocity shift of the Maxwellian. Because $f(v_0)$
is derived from the fit of Jacobian to the observed brightness, it does not
refer to the absolute fraction of jet particles with an initial velocity $v_0$.
However, since brightness is proportional to electron number density, $f(v_0)$
is thus proportional to the normalized velocity distribution. Therefore, we added
a coefficient $C$ in Equ.~\ref{equ:maxw1d}.
In the left panel of Fig.~\ref{fig:f0v0}, the parameters $\epsilon_k$
and $v_b$ are chosen to obtain the best fit to the observed values of
$f(v_0)$. The fit results in the values of
$\epsilon_k = 2.6\times10^{-10}$ erg for the
mean kinetic energy and $v_b =$ 230 km\,s$^{-1}$ for the velocity shift. 

Because orbits with an initial velocity below 400 km\,s$^{-1}$ do
not reach the field of view of COR1 and COR2,
we are lacking $f(v_0)$ in the velocity range below this threshold.
For this reason, our fit is somewhat insensitive in particular to $v_b$.
We therefore also considered another fit in the right panel where $v_b$ is
set to zero. We then found a higher mean kinetic 
energy of $5.4\times10^{-10}$ erg.

In either case, we found an enhanced higher energy tail at $v_0 \geq 600$
km\,s$^{-1}$ that cannot be fitted to a Maxwellian. This deviation from
a Maxwellian in the jet source distribution may be the result of a specific
heating and acceleration process of the jet particles.

In addition to the jet initiation time and the initial velocity distribution we have
determined the absolute mass ejected by the jet. For a white light coronagraph
image, the electron column density scattering into each pixel is proportional
to the calibrated pixel total brightness.
Since we have subtracted the
white-light background, only the jet particle density remains from this
calculation. Moreover, the brightness in Fig.~\ref{fig:DT_euco} was also
integrated across the jet widths.
If we additionally integrate along the distance $s$, we obtain the total
number of jet particles visible in the coronagraphs at any one time.

The relations between the image brightness $B$ and the electron density $N_e$
were established by \citet{Minnaert:1930}, \citet{VandeHulst:1950} and
\citet{Billings:1966}. For our purposes, we modified them to
\begin{equation}
B=\frac{B_\odot}{1-u/3}\frac{\pi\sigma N_e dl}{2}[(1-u)(2C-A\sin^2\chi)]+
u(2D-B\sin^2\chi)],
\label{equ:totB}
\end{equation}
where
$B_\odot$ is the physical mean solar brightness (MSB) that is used as units
in which calibrated COR1 and COR2 observations are expressed.
$A,B,C,and D$ are known functions of the distance of the scattering location
from the Sun's center that express the dependence of the scattered
polarization on the size of the solar disk as seen from the scatterer.
$u$ accounts for the solar disk limb darkening and
$\sigma$ is the differential Thomson scattering cross section.
For $u$, we used the conventional value of 0.56 for the white-light spectral
range.
Finally, $N_e\,dl$ is the electron column density along the line-of-sight
direction through the jet.
As an approximation, we assumed that the jet was lying in the plane of sky, so
that we have a scattering angle $\chi=90^\circ$ for the Thomson scattering.

For the jet observations at $t=$ 05:45 UT when it was most 
prominent, the total brightness has been integrated across the width, say 
integrated over $dq$ to derive $B(s,t-t_0)$, which again was integrated over $ds$ 
along the jet. If we relate this final integration to Equ.~\ref{equ:totB}, we
derive $\int N_e dl dq ds$. The result corresponds to a total number of
$1.1\times 10^{37}$ jet particles seen above the occulter at this time instance.

From our trajectory analysis this number is related to the particles in the
velocity range of $v_0$ from 400 to 600 km\,s$^{-1}$. Depending on the
extrapolation in velocity space presented in Fig.~\ref{fig:f0v0}, we may
extend the above particle number estimate to the total number of jet
particles. The two cases shown there can be considered as two extreme cases.
The total number of jet particles obtained this way will then lie between
$1.6\times 10^{38}$ (Fig.~\ref{fig:f0v0} left case) and $8.9\times 10^{38}$
(Fig.~\ref{fig:f0v0} right case). It corresponds to the
jet mass between $3.2\times10^{14}$ and $1.8\times10^{15}
$ g.
Note that these extrapolations must be treated with some care. They rely on
the fact that the distribution has a strictly Maxwellian core and even
the respective parameters are estimates only based on the observed
distribution of the far tail. Also, the resulting particle number
may seem large compared to previous estimates merely from coronagraph 
observations, e.g. the COR 1 data at 05:45 UT, because it includes all 
particles involved, also those with a low initial velocity $v_0$,
which are not seen in the coronagraph images at all.

We may use these number estimates to speculate about
the particle number density in the source region. If
we divide the particle number by a typical source volume $V_{BP}$, we
obtain the probable number density at the height where the jet
heating and/or acceleration took place.
Here, the volume of this jet source region is assumed to be
the apparent size of the bright point in the 195~\AA~pre-event image.
We estimate this volume to $2\times10^{28}$ cm$^{3}$, which yields an electron
density between $8\times10^9$ cm$^{-3}$ for the case that the jet was
heated and accelerated, and $5\times10^{10}$ cm$^{-3}$ for the case that
the jet was only heated.
The densities are typical for the upper chromosphere or the low corona and
we may conclude that the jet material was heated and possibly accelerated
in this height region.
Based on the values for $N_e$, $\epsilon_k$ and $V_{BP}$ above,  
the total kinetic energy of all jet particles in the source region
 $E_k=(1/2)N_e\epsilon_kV_{BP}$ can be estimated
which was between $2.1\times10^{28}$ erg and
$2.4\times10^{29}$ erg. This energy is consistent with typical
energies for microflares, indeed our estimate lies in the higher energy
range of the microflare energy spectrum from $10^{27}$ to 
$10^{30}$ erg. Again we caution the reader that these numbers
are based on the extrapolation of the particle velocity distribution as shown
in Fig~\ref{fig:f0v0}. In particular, we implicitly use $v_0 =$ 0 km\,s$^{-1}$
as the lower bound of the distribution for jet particles. The uncertainty
of this lower bound has a strong impact on the estimated total number of 
jet particles, less impact on the kinetic energy in the source region.

\subsection{Mirror force and collisions}
\label{sec:mirror+coll}

The fits of $J(v_0,t)$ to $f(v_0)/B(s,t)$ in Fig.~\ref{fig:jf0Berr} as
functions of time $t$ after jet initiation are in general good but not
perfect. A good deal of this imperfection can be attributed to image noise,
especially for high values of $t$, i.e., low observed brightnesses $B(s,t)$.
This could be considered as evidence that the particle orbit is sufficiently
well described by the action of the gravity force alone.
In this subsection we discuss the terms that we neglected from the full
equation (\ref{equ:kinem1}), the mirror force and collisions, and find out
the conditions under which they might become important.

We approximate the Sun's polar magnetic field by a dipole field. Then the
magnetic field strength above the pole is
\begin{equation}
H=H_\odot\frac{r^3_\odot}{r^3}\;,\quad
\frac{\partial{H}}{\partial{r}}=-3H_\odot\frac{r^3_\odot}{r^4}\;,\quad
\end{equation}
where $H_\odot$ is the field strength at the pole.
The mirror force on an individual particle can then be written as
$F_m=\mu\partial{H}/\partial{r}$ and therefore decreases as $\propto r^{-4}$
with distance from the solar surface. Note that $\mu=mv_\perp^2/2H$ is the
particle's magnetic moment and an adiabatic invariant of motion provided that
the spatial variations of the magnetic field are smooth.
On the other hand, the gravity felt by the particle decreases as
$\propto r^{-2}$ and hence less rapidly than the mirror force.
On the surface of the Sun, comparison of the mirror force and gravity force
shows that the latter dominates as long as the local
escape velocity $(2 g_{\odot}r_{\odot}^2/r(s))^{1/2}$ well exceeds the
perpendicular velocity $v_\perp$. For a thermal speed on the order of
the escape velocity of 618 km\,s$^{-1}$, a temperature of 46~MK is required.

The other effect neglected in our analysis are possible collisions of the
jet particles with the coronal plasma background. In order to estimate
this effect, we used as an estimate of the collisional deceleration 
$\mathrm{a}_\mathrm{coll}$ the friction coefficient of the Fokker-Planck
collision term in a kinetic plasma description(e.g., \citet{Ishimaru:1973}).
The deceleration depends on the velocity $v$ of the jet particle
relative to the thermal velocity $v_\mathrm{therm}$ of the coronal
background, which is assumed at rest. Then for singly charged particles,
\begin{equation}
a_\mathrm{coll}(v) = -\mathrm{sign}(v) \; \frac{e^2}{m \lambda_D^2}
                      \ln\Lambda \; G(\frac{v}{v_\mathrm{therm}}),
\label{a_coll}\end{equation}
where $e$ is the electron charge, $\lambda_D$ the coronal plasma Debye length
and $\ln\Lambda$ the respective Coulomb logarithm. An implicit assumption in
Equ.~\ref{a_coll} is that the background plasma has a Maxwellian velocity
distribution. The velocity dependence in the Fokker-Planck collision term
generally expressed by Rosenbluth potentials can then be reduced to the
Chandrasekhar function \citep{Rosenbluth:etal:1957} 
\[
 G(x)=\frac{1}{2}\frac{d}{dx}(\frac{\mathrm{erf}(x)}{x})
\]
The velocity of the jet particles is well ahead of $v_\mathrm{therm}$ on a
large part of their orbit, therefore $G$ is needed for large arguments for
which it decays as $\simeq 1/2x^2$. Hence faster particles feel less
collisional deceleration, which eventually could lead to a runaway of
energetic particles (e.g., \citet{Dreicer:1959,Springmann:Pauldrach:1992}). 

\begin{figure} % fig 9
  \centering
  %\hbox{
  \vbox{
  \includegraphics[bb=70 40 385 360, clip=true, height=6.cm]{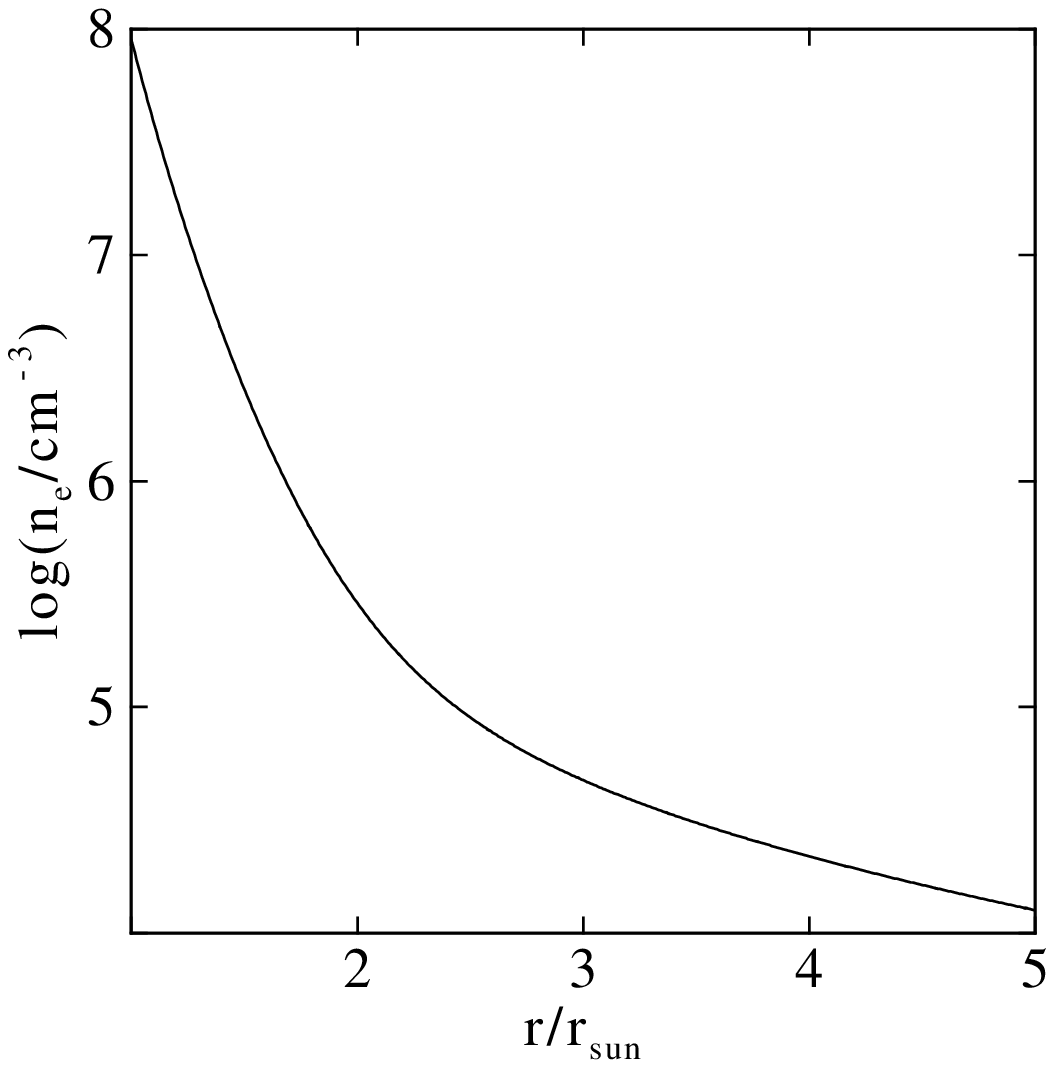}
  \includegraphics[bb=30 40 400 360, clip=true, height=6.cm]{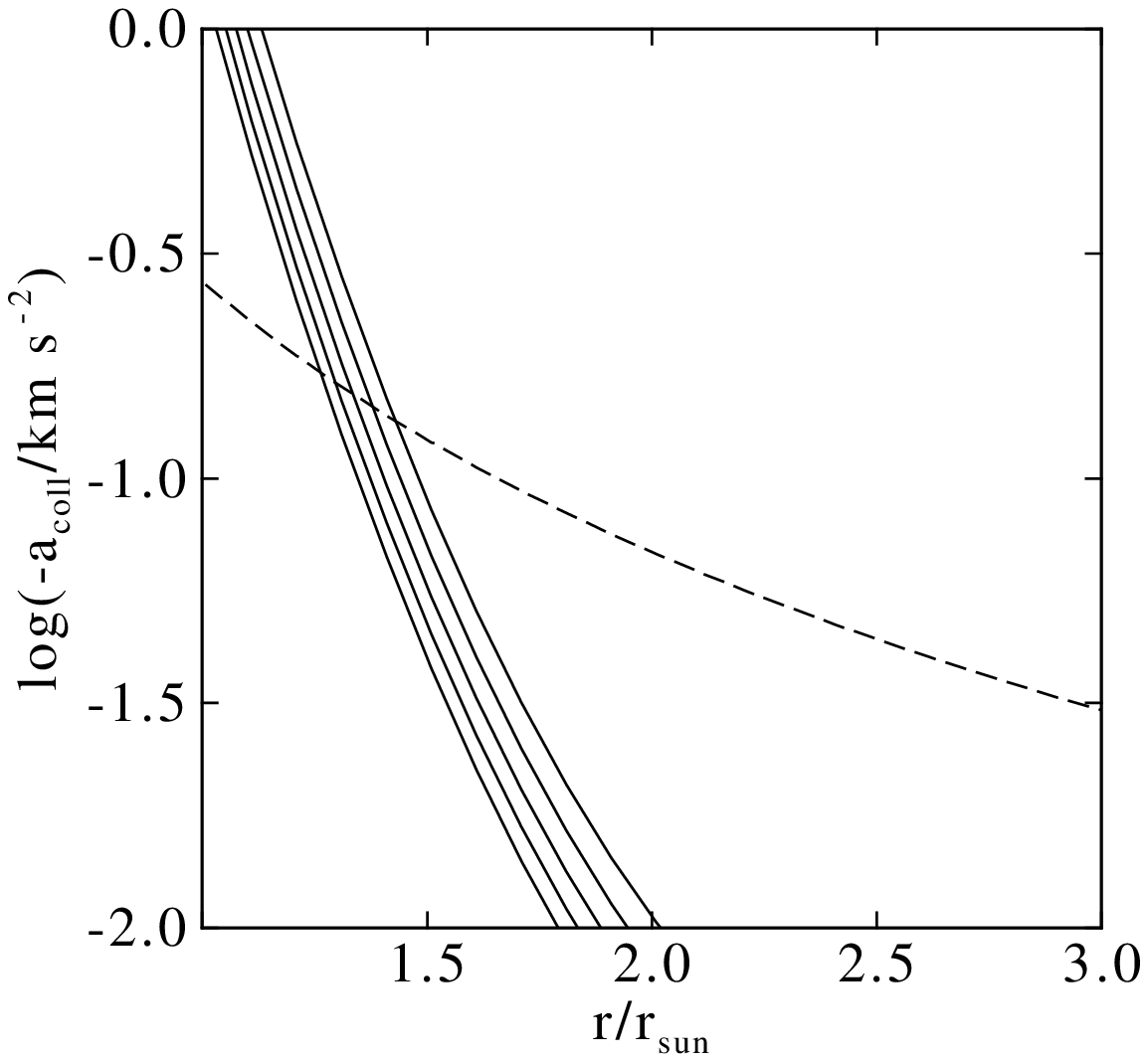}}
  \caption{Top: Electron density assumed for the background corona,
  adapted from \citet{Quemerais:Lamy:2002}. Bottom: Collisional deceleration
  of the jet particles for velocities of 400 to 600 km\,s$^{-1}$
  in steps of 50 km\,s$^{-1}$ (solid, from right to left) compared to
  the gravitational accelaration (dashed).}
  \label{fig:collision}
\end{figure}

For our estimates we assumed a density as in
the left panel of Fig.~\ref{fig:collision} adopted
from \citet{Quemerais:Lamy:2002}.
In the right panel of Fig.~\ref{fig:collision} the resulting
deceleration $a_\mathrm{coll}$ for different velocities and the gravitational
acceleration are compared. The neglect of collisions in our analysis is justified above
about 1.4 $r_\odot$ for particles with 400 km\,s$^{-1}$ 
and for even lower heights for faster particles. For heights below, the
coupling of the jet particles to the background plasma is very intense
and a single particle analysis of the jet seems no longer justified.

The effect these collisions will probably have is that part of the jet's momentum
will be transferred to the background plasma, which in turn is accelerated.
Therefore not all particles we see above the occulter may be original jet
particles, and the distribution function we deduced in Fig.~\ref{fig:f0v0} may
be the result of some interaction of the original jet with the background
plasma. The fact that the distribution in Fig.~\ref{fig:f0v0} closely fits to
a Maxwellian at lower velocities may be evidence for these collisional
processes at lower heights.

In the above collision estimate, we have ignored the solar wind in the
distribution of the coronal background particles. We expect that
this does not alter our conclusions substantially because below
about 1.4 $R_\odot$, where collisions matter, the solar wind is probably
still subsonic. At greater heights, where the solar wind speed becomes
significant, the collisional coupling of the jet particles is scarce.

However, a more serious point is that the acceleration mechanism that
drives the solar wind may also affect the jet particles and should then
be added as additional term to the right-hand side of Equ.~\ref{equ:kinem1}. A
physical mechanism for this acceleration has, however, not been identified
yet and any such acceleration term would be highly speculative.

\section{Conclusions}

We have followed the evolution of a big eruptive jet event
observed by SECCHI in both EUVI images and the COR1 and COR2 white-light
coronagraphs. Based on the distance-time brightness analysis,
we found that a ballistic model for the jet particles in general can
explain quite well the brightness variation beyond 1.5 $r_\odot$ in the COR1
and COR2 fields of view with gravity as the dominant acceleration of
the jet particles.

Additional parameters were derived or at least estimated, such as the initiation
time, the initial velocity distribution, and the number of the jet particles.
The derived initiation time is consistent with the EUVI
observations at lower altitudes. The initial velocity distribution was 
fitted by two Maxwellian distributions with different mean kinetic energies.
The good agreement with a Maxwellian for
lower initial velocities may be due to collisions at heights below 1.4
$r_\odot$. At high initial velocities, the distribution deviates from the
Maxwellian toward a power law tail that may be a result of the jet
acceleration process. The total jet particle number and kinetic energy sum up
to about 1.6 to 8.9$\times10^{38}$ and 2.1 to 24$\times10^{28}$ erg, respectively.

We neglected the effect of the magnetic mirror force and of Coulomb
collision. As discussed, they might have some effect on the
kinetics of the jet particles at lower altitudes. Note that these two forces
counteract each other: while the collisions with the background plasma will
decelerate the particles, the mirror force accelerates them away from the Sun.
Especially the correct modeling of the Coulomb collisions below 1.4 $r_\odot$
requires additional assumptions, e.g., about the coronal background density
and its velocity.

We outlined the basic idea of a new kinetic jet analysis.
In the future, a more sophisticated kinetic model of the jet may be compared to
white-light observations. We have shown that this comparison allows one to
constrain details of the jet which could not be derived in previous studies.
More work needs to be done in these directions. Moreover, more jet
samples are required to find out to which extent a jet is embedded in the
ambient solar wind and how the jet interacts with it.

PROBA-3, which will be launched in a few years, will have a coronagraph with a
FOV of 1.04 to 3 $r_\odot$. It will provide us with a broader initial velocity
coverage because the lower velocity limit depends on the occulter's size .
Therefore we will have less uncertainty in the initial velocity distribution,
the electron density in the jet source region, etc. The higher temporal
observations with more wavelength coverage from AIA/SDO will help determine
the jet initiation time more precisely.

Recently, \citet{Raouafi:etal:2008} found that a jet was very often succeeded
by a plume above the jet launch site. Interestingly, in our jet
study a plume was visible in both EUVI and COR1 before the jet. This phenomenon
was also observed by \citet{Lites:etal:1999} and other white-light
observations in LASCO C2 \citep{Llebaria:etal:2002}. However, no definite
conclusion is given concerning the relation between plume and jet. A time
series of 3D reconstruction of both plume and jet needs to be made to find the
answer to this question.

\acknowledgements

The authors thank the referee for his/her detailed and constructive 
comments which have improved the work shown in this manuscript.  
STEREO is a project of NASA. The SECCHI data used here were produced by an
international consortium of the Naval Research Laboratory (USA),
Lockheed Martin Solar and Astrophysics Lab (USA), NASA Goddard Space
Flight Center (USA), Rutherford Appleton Laboratory (UK), University of
Birmingham (UK), Max-Planck-Institut for Solar System Research (Germany),
Centre Spatiale de Li\`ege (Belgium), Institut d'Optique Th\'eorique et
Applique\'e (France), Institut d'Astrophysique Spatiale (France).
The work was supported by DLR grant 50 OC 0904, NSFC grant 11003047 and
973 Program 2011CB811402.


\begin{thebibliography}{}

\bibitem[Billings(1966)]{Billings:1966} Billings, D.~E.\ 1966, New
York: Academic Press, 1966

\bibitem[Domingo et al.(1995)]{Domingo:etal:1995} Domingo, V., Fleck, B.,
\& Poland, A.~I.\ 1995, \solphys, 162, 1

\bibitem[Dreicer(1959)]{Dreicer:1959} Dreicer, H.\ 1959, Physical Review,
  115, 238
  
\bibitem[Howard et al.(2008)]{Howard:etal:2008} Howard, R.~A., et al.\
2008, Space Science Reviews, 136, 67

\bibitem[Ishimaru (1973)]{Ishimaru:1973} Ishimaru, S., \
1973, Basic principles of Plasma Physics, Benjamin Inc., London

\bibitem[Kamio et al.(2010)]{Kamio:etal:2010} Kamio, S., Curdt, W.,
Teriaca, L., Inhester, B., \& Solanki, S.~K.\ 2010, \aap, 510, L1


\bibitem[Ko et al.(2005)]{Ko:etal:2005} Ko, Y.-K., et al.\ 2005,
\apj, 623, 519

\bibitem[Llebaria et al.(2002)]{Llebaria:etal:2002} Llebaria, A.,
Thernisien, A., \& Lamy, P.\ 2002, Advances in Space Research, 29, 343


\bibitem[Lites et al.(1999)]{Lites:etal:1999} Lites, B.~W., Card, G.,
Elmore, D.~F., Holzer, T., Lecinski, A., Streander, K.~V., Tomczyk, S.,
\& Gurman, J.~B.\ 1999, \solphys, 190, 185


\bibitem[Minnaert(1930)]{Minnaert:1930} Minnaert, M.\ 1930, \zap, 1,209


\bibitem[Moses et al.(1997)]{Moses:etal:1997} Moses, D., et al.\ 1997,
\solphys, 175, 571


\bibitem[Patsourakos et al.(2008)]{Patsourakos:etal:2008} Patsourakos, S.,
Pariat, E., Vourlidas, A., Antiochos, S.~K.,
\& Wuelser, J.~P.\ 2008, \apjl, 680, L73


\bibitem[Qu{\'e}merais \& Lamy(2002)]{Quemerais:Lamy:2002}
Qu{\'e}merais, E., \& Lamy, P.\ 2002, \aap, 393, 295


\bibitem[Raouafi et al.(2008)]{Raouafi:etal:2008} Raouafi, N.-E., Petrie,
G.~J.~D., Norton, A.~A., Henney, C.~J.,
\& Solanki, S.~K.\ 2008, \apjl, 682, L137


\bibitem[Rosenbluth et al. (1957)]{Rosenbluth:etal:1957}
  Rosenbluth, M.~N. and MacDonald, W.~M. and Judd, D.~L.\
  1957, Physical Review, 115, 238
  
\bibitem[Shibata et al.(1992)]{Shibata:etal:1992} Shibata, K., et al.\
1992, \pasj, 44, L173


\bibitem[Shibata et al.(2007)]{Shibata:etal:2007} Shibata, K., et al.\
2007, Science, 318, 1591

\bibitem[Springmann and Pauldrach (1992)]{Springmann:Pauldrach:1992}
  {Springmann}, U.~W.~E. and {Pauldrach}, A.~W.~A.\
1992, {\aap}, 262, 515

\bibitem[St.~Cyr et al.(1997)]{StCyr:etal:1997} St.~Cyr, O.~C., et al.\
1997, Correlated Phenomena at the Sun, in the Heliosphere and in Geospace,
415, 103

\bibitem[van de Hulst(1950)]{VandeHulst:1950} van de Hulst, H.~C.\
1950, \bain, 11, 135

\bibitem[Wang et al.(1998)]{Wang:etal:1998} Wang, Y.-M., et al.\ 1998,
\apj, 508, 899

\bibitem[Wilhelm et al.(2011)]{Wilhelm:etal:2011} Wilhelm, K., et al.\
2011, The Astronomy and Astrophysics Review, 19, 35

\bibitem[Wood et al.(1999)]{Wood:etal:1999} Wood, B.~E., Karovska, M.,
Cook, J.~W., Howard, R.~A., \& Brueckner, G.~E.\ 1999, \apj, 523, 444


  


%%@Article{PhysRev.115.238,
%%  title = {Electron and Ion Runaway in a Fully Ionized Gas. I},
%%  author = {Dreicer, H. },
%%  journal = {Phys. Rev.},
%%  volume = {115},
%%  number = {2},
%%  pages = {238--249},
%%  numpages = {11},
%%  year = {1959},
%%  month = {Jul},
%%  doi = {10.1103/PhysRev.115.238},
%%  publisher = {American Physical Society}
%%}
%%@Article{PhysRev.117.329,
%%  title = {Electron and Ion Runaway in a Fully Ionized Gas. II},
%%  author = {Dreicer, H. },
%%  journal = {Phys. Rev.},
%%  volume = {117},
%%  number = {2},
%%  pages = {329--342},
%%  numpages = {13},
%%  year = {1960},
%%  month = {Jan},
%%  doi = {10.1103/PhysRev.117.329},
%%  publisher = {American Physical Society}
%%}

\end{thebibliography}
\end{document}